\documentclass[prd,onecolumn,nofootinbib,showpacs]{revtex4}
\usepackage{epsfig,graphicx,bm}
\usepackage[latin2]{inputenc}
\usepackage{color}
\usepackage{t1enc}                                                             
\usepackage{amssymb}
\usepackage{amsmath}

\newcommand{\tr}{\textnormal{Tr\,}}
\newcommand{\bea}{\begin{eqnarray}}
\newcommand{\eea}{\end{eqnarray}}
\newcommand{\be}{\begin{equation}}
\newcommand{\ee}{\end{equation}}

\def\lag{\langle}
\def\rag{\rangle}
\newcommand{\no}{{\nonumber}}

\newcommand{\rt}[1]{{}}

\begin{document}

\title{Mapping the boundary of the first order
finite temperature  restoration of chiral symmetry
in the $(m_\pi - m_K)$--plane with a linear sigma model}

\author{T. Herpay}
\email{herpay@complex.elte.hu}
\affiliation{Department of Physics of Complex Systems,
E{\"o}tv{\"o}s University, H-1117 Budapest, Hungary}
\author{A. Patk{\'o}s}
\email{patkos@ludens.elte.hu}
\affiliation{Department of Atomic Physics, E{\"o}tv{\"o}s University,
H-1117 Budapest, Hungary}
\author{Zs. Sz{\'e}p}
\email{szepzs@achilles.elte.hu}
\affiliation{Research Group for Statistical Physics of the
  Hungarian Academy of Sciences, H-1117 Budapest, Hungary}
\author{P. Sz{\'e}pfalusy}
\email{psz@galahad.elte.hu}
\affiliation{Department of Physics of Complex Systems,
E{\"o}tv{\"o}s University, H-1117 Budapest, Hungary}
\affiliation{Research Institute for Solid State Physics and Optics,
Hungarian Academy of Sciences, H-1525 Budapest, Hungary}

\begin{abstract}
The phase diagram of the three-flavor QCD is mapped out in the low
mass corner of the $(m_\pi~-~m_K)$--plane with help of the
$SU_L(3)~\times~SU_R(3)$ linear sigma model (L$\sigma$M). A novel zero
temperature parametrization is proposed for the mass dependence of the
couplings away from the physical point based on the the three-flavor
chiral perturbation theory (U(3) ChPT).  One-loop thermodynamics is
constructed by applying optimized perturbation theory. The unknown
dependence of the scalar spectra on the pseudoscalar masses limitates
the accuracy of the predictions. Results are compared to lattice data
and similar investigations with other variants of effective chiral
models. The critical value of the pion mass is below $65$ MeV for all
$m_K$ values $\lesssim 800$ MeV. Along the diagonal $m_\pi=m_K$, we
estimate $m_\textrm{crit}(\textrm{diag})=40\pm 20$ MeV.
\end{abstract}

\maketitle

\section{Introduction}

The ambition of the exploration of the QCD phase structure
corresponding to different breaking patterns of $SU_L(3)~\times~SU_R(3)$ 
chiral symmetry is the determination of the true ground state
of the theory for an arbitrary set of quark masses $m_u, m_d, m_s$ in
presence of a variety of intensive thermodynamical parameters,
e.g. temperature ($T$), baryonic ($\mu_B$), isospin ($\mu_I$) and
strangeness ($\mu_s$) chemical potentials.  The progress is continuous
both in numerical lattice simulations \cite{petreczky04} and in the
application of effective models \cite{barducci05,roder03,toublan03,Lianyi}
for extracting results of phenomenological interest.  The baryonic
density of the Early Universe was very small when the cosmic expansion
drove it through the stages of chiral symmetry breaking (the
condensation of the different quark flavors). Also for the extreme
high energies of heavy ion collisions achieved at RHIC the average
baryonic density of the final state is very close to zero. This
motivates the present investigation where we concentrate on the case
when all types of chemical potential vanish.

Universality arguments \cite{pisarski84} predict first order
transition for $m_u=m_d=m_s=0$ and a second order one for $m_u=m_d=0,
m_s=\infty$.  One expects the existence of a triple point for some
$m_s=m_{s,c}$.  The most systematic effort seeking the explicit
solution of the thermodynamics of the 3-flavor QCD is done with help
of numerical simulations in the bulk of the $(m_u=m_d, m_s)$-plane
\cite{karsch03,bernard04}. However, by the nature of the lattice
regularization, one explored to date mostly the region of rather
massive $u-d$ quarks, usually corresponding to pion masses of order
3-500 MeV (in these simulations $m_s$ is mostly kept fixed at
its physical value). Lattice version of chiral perturbation theory
(ChPT) is employed for extrapolating the results to the physical mass
point. Also finite lattice spacing effects turned out rather
important, therefore improved lattice actions gained significance in
reaching physical conclusions.  Common wisdom at present concludes
that in the physical point temperature variations move the
thermodynamical potential of the system analytically between the chirally
symmetric and the broken symmetry regimes.

At the same time constant interest is manifested concerning the
location of the borderline of the region of first order
transitions. If the border passes nearby, one might expect it to
influence in a substantial way the transformation of the physical
ground state \cite{gavin94,hatta03}.  Numerical investigations were
done and systematically improved for the 3-flavor degenerate case
$m_u=m_d=m_s\neq 0$. The initial estimate of
$m_\textrm{crit}(\textrm{diag})\approx 290$ MeV \cite{karsch01} was
seen to be reduced to $60-70$ MeV \cite{karsch03} or may be to even
further down \cite{bernard04} when finer lattices and improved lattice
actions are used.

Effective models (linear or nonlinear sigma models,
Nambu--Jona-Lasinio model) represent another, in a sense
complementary, approach to the study of the phase structure, which one
expects to work the better the lighter quark masses are used
\cite{Meyer-Ortmanns,Lenaghan00}. It is surprising that only moderate
effort was invested to date to improve the pioneering studies of the
$SU(3)~\times~SU(3)$ linear sigma model by Meyer-Ortmanns and Schaefer
\cite{Meyer-Ortmanns} who used a saddle point approximation valid in
the limit of infinite number of flavors, and derived
$m_\textrm{crit}(\textrm{diag})\lesssim 51$ MeV. An extension of their
work to unequal pion and kaon masses was achieved by C. Schmidt
\cite{schmidt03}. He found $m_\textrm{crit}(\textrm{diag})=47$ MeV and a 
phase boundary approaching the $m_K$-axis rather sharply. The location of
the tricritical point can be estimated from extrapolating his curve to
$m_{K}(\textrm{tricrit})\approx 70$ MeV, although the expected
power-scaling of the boundary curve with $m_\pi$ is difficult to
disentangle from a simple linear regime.  The phase boundary was
calculated also by Lenaghan \cite{lenaghan01} using the
Hartree-approximation to the effective potential derived in CJT-formalism 
\cite{CJT}. For the complete determination of the couplings of the
three-flavor chiral meson model he fixed the $T=0$ mass of the
$\sigma$ particle in addition to the phenomenology of the pseudoscalar
sector. The emerging phase boundary is rather sensitive to this mass.
For instance in the case of $U_A(1)$ anomaly, the tricritical kaon
mass is $m_K(\textrm{tricrit})\approx 161$ MeV ($m_s=16$ MeV) for
$m_\sigma=800$ MeV, and the expected $m_\pi$ scaling is not seen,
while for $m_\sigma=900$ MeV, $m_K(\textrm{tricrit})\approx 652$ MeV
($m_s=260$ MeV).  The estimate for $m_\textrm{crit}(\textrm{diag})$
which one can extract from Fig.~3 of \cite{lenaghan01} for
$m_\sigma=900$ MeV is compatible with \cite{Meyer-Ortmanns,schmidt03}.
 
In our opinion the greatest problem in refining the linear sigma model into
a competitive tool of investigation of the chiral phase diagram is the
difficulty of the determination of the quark (pseudoscalar meson) 
mass dependence of the couplings of the effective models. Almost all
investigations tune exclusively the strength of explicit chiral symmetry
breaking to cope with the variation of the pion and kaon masses via the
Gell-Mann--Oakes--Renner relation. All other couplings are usually kept at
the values determined in the physical point. One might note, however, some
attempts to include also the variation of $f_\pi$ as deduced from lattice
studies \cite{Lenaghan00}.

The novel feature of our paper is the parametrization of the couplings
of the 3-flavor linear sigma model which ensures a full agreement with
the results of ChPT for the variation of the tree--level pseudoscalar
mass spectra as a function of the pion and kaon masses.  In order to
make the paper self-contained, we review in Section 2 the
parametrization of the linear sigma model, which essentially follows
Refs.~\cite{Tornqvist99,Lenaghan00}. In Section 3 the relevant
$\mathcal{O}(1/f^2)$ accurate results of ChPT
\cite{Gasser85,Herrera98,Herrera97,Borasoy01,Beisert01} are summarized
and used for the determination of the ($m_\pi, m_K$)--dependence of
the L$\sigma$M--couplings.  Full details of the parametrization can be
reproduced with help of three Appendices.  Next, we derive in Section
4 the equations of state for the nonstrange and strange condensates
together with the gap equation for the common thermal mass which
characterizes the finite temperature behavior of the scalar and
pseudoscalar spectra. For this we use a variant of the Optimized
Perturbation Theory \cite{chiku}. In this way we partially avoid the
imaginary mass problem of the standard loop expansions emphasized by
\cite{Lenaghan00}. In section 5 we argue that the phase boundary
separating the region of first order transitions from the crossover
regime varies sensitively depending on the assumption we make about
the scalar sector when specifying the couplings of the model. Inspite
of this variation we are able to conclude that the critical pion mass
does not exceed $65$ MeV in the region $0<m_K<800$ MeV. In particular
the $m_\pi=m_K$ diagonal is crossed by the phase boundary in the
region $20\ \textrm{MeV}< m_\textrm{crit}(\textrm{diag})<65$ MeV.

\section{Tree level parametrization of the couplings}

\noindent The Lagrangian of the $SU_L(3)~\times~SU_R(3)$ symmetric 
linear sigma model 
with explicit symmetry breaking terms is given by \cite{Haymaker73}
\begin{equation}
L(M)=\frac{1}{2}\tr(\partial_\mu M^{\dag} \partial^\mu M+\mu_0^2
M^{\dag} M)-f_1
\left( \tr(M^{\dag} M)\right)^2-f_2  \tr(M^{\dag}
M)^2-g\left(\det(M)+\det(M^{\dag})\right)+\epsilon_0\sigma_0+\epsilon_8 
\sigma_8,
\label{Lagrangian}
\end{equation}
where $M$ is a complex 3$\times$3 matrix, defined by the $\sigma_i$ scalar
and $\pi_i$ pseudoscalar fields 
$\displaystyle M:=\frac{1}{\sqrt{2}}\sum_{i=0}^{8}(\sigma_i+i\pi_i)\lambda_i$,
with  $\lambda_i\,:\,\,\,i=1\ldots 8$ the Gell-Mann matrices and  
$\lambda_0:=\sqrt{\frac{2}{3}} {\bf 1}.$
The last two terms of (\ref{Lagrangian}) break the symmetry
explicitly, the possible isospin breaking term $\epsilon_3\sigma_3$ is
not considered. 

A detailed analysis of the symmetry breaking patterns
which might occur in the system described by this Lagrangian can be
found in \cite{Lenaghan00}. The fields $\sigma_0, \sigma_8$ both
contain strange and nonstrange components. For the purpose of the
exploration of the $(m_\pi - m_K)$-dependence of the phase diagram we
found more convenient to decompose the vacuum condensate into strange
and nonstrange parts which is realized by an orthogonal
transformation in the algebra basis and also defined the corresponding
external fields:
\be
\begin{pmatrix}\sigma_x\\ \sigma_y\end{pmatrix}:=O
\begin{pmatrix}\sigma_0\\ \sigma_8\end{pmatrix},
\begin{pmatrix}\pi_x\\ \pi_y\end{pmatrix}:=O\begin{pmatrix}\pi_0\\ \pi_8
\end{pmatrix}, \qquad
\begin{pmatrix}\epsilon_x\\ \epsilon_y\end{pmatrix}:=O
\begin{pmatrix}\epsilon_0\\ \epsilon_8\end{pmatrix},
\ee
where 
\be
O:=\frac{1}{\sqrt{3}} 
\begin{pmatrix}  \sqrt{2}   & \,\quad  1  \\  1\,     & -\sqrt{2}
\end{pmatrix} \,.
\label{Otrans}
\ee
The fields with indices $x, y$ appear in the matrix $M$ as follows 
\begin{equation}
M=
\frac{1}{\sqrt{2}}\sum_{i=1}^{7}(\sigma_i+i\pi_i)\lambda_i+\frac{1}{\sqrt{2}}
\textrm{diag}(\sigma_x+i\pi_x,\sigma_x+i\pi_x,\sqrt{2}(\sigma_y+i\pi_y))
.
\label{xybasis}
\end{equation}

For the tree--level determination of the parameters of the system we have at
our disposal the equations of state, the mass spectra of the pseudoscalar
and scalar nonets and the consequences of Partially Conserved Axial-Vector 
Current (PCAC) relations  for the weak decay of $\pi$
and $K$.  After some algebra (cf. \cite{Haymaker73}) one obtains the zeroth
order term of the Lagrangian in the fluctuations around the expectation
values $<\sigma_x>=:x\,,<\sigma_y>=:y$, which is the classical potential
\begin{equation} 
U_{cl}=-L\Big|_{\scriptsize
\begin{gathered} \sigma_x=x\vspace*{-0.18cm}\\ \sigma_y=y
\end{gathered}}=-\epsilon_x
x-\epsilon_y y-\frac{\mu_0^2}{2}(x^2+y^2)+gx^2y+2f_1 x^2
y^2+(f_1+\frac{ f_2}{2})x^4+
(f_1+ f_2)y^4.
 \label{ucl}
\end{equation}
The terms linear in the fluctuations must vanish, accordingly the two
equations of state are
\begin{eqnarray}
E_x:=\left.\frac{\partial L}{\partial
  \sigma_x}\right|_{\scriptsize
\begin{gathered} \sigma_x=x\vspace*{-0.18cm}\\ \sigma_y=y \end{gathered}
} &= & \epsilon_x+\mu_0^2
x-2gxy-4f_1xy^2-2(2f_1+f_2)x^3=0, 
\label{eq:ex}\\
E_y:=\left.\frac{\partial L}{\partial
  \sigma_y}\right|_{\scriptsize
\begin{gathered} \sigma_x=x\vspace*{-0.18cm}\\ \sigma_y=y \end{gathered}
} &= & \epsilon_y+\mu_0^2
y-gx^2-4f_1x^2y-4(f_1+f_2)y^3=0 \label{eq:ey}
\textnormal{ .}
\end{eqnarray} 
The matrix of the squared masses can be read from the coefficients of the
quadratic terms, see Table~\ref{Tab:masses}. 
There is a mixing in the $x-y$ sector
represented by entries of the last three rows.  The mass matrix of $\eta$
fields is given also in the  $\eta_{0} -\eta_{8}$ basis in 
Appendix \ref{App:eta_08}. The third and fourth order terms
yield the three-- and 
four--point interaction vertices. Finally, PCAC relates the condensates $x$ and
$y$ to the pion ($f_\pi$) and kaon ($f_K$) decay constants:
\begin{table}
\label{Tab:masses}
\begin{tabular}{|l|l|}
\hline
$m^2_\pi\,\,\,\,\,=-\mu_0^2+2(2f_1+f_2)x^2+4f_1y^2+2gy$ &
$m^2_{a_0}\,\,=-\mu_0^2+2(2f_1+3f_2)x^2+4f_1y^2-2gy$ \\\hline
$m^2_K\,\,\,=-\mu_0^2+2(2f_1+f_2)(x^2+y^2)+2f_2y^2-\sqrt{2}x(2f_2y-g)$
&$
 m^2_{\kappa}\,\,\,\,\,=-\mu_0^2+2(2f_1+f_2)(x^2+y^2)+2f_2y^2+
\sqrt{2}x(2f_2y-g)$ 
\\\hline
$m^2_{\eta_{xx}}=-\mu_0^2+2(2f_1+f_2)x^2+4f_1y^2-2gy$ & $
m^2_{\sigma_{xx}}=-
\mu_0^2+6(2f_1+f_2)x^2+4f_1y^2+2gy$ \\\hline
$ m^2_{\eta_{yy}}=-\mu_0^2+4f_1x^2+4(f_1+f_2)y^2$ & $
m^2_{\sigma_{yy}}=-
\mu_0^2+4f_1x^2+12(f_1+f_2)y^2$ \\\hline
$m^2_{\eta_{xy}}=-2gx$ &$ m^2_{\sigma_{xy}}=8f_1xy+2gx $ \\\hline
\end{tabular}
\\
\caption{
The squared masses of the pseudoscalar nonet appear in the first column. The
first two entries are the squared masses of pions and kaons, the last three
rows represent the mixing in the $\eta - \eta'$ sector. The second column
contains the same quantities for the scalar parity partners.  The
phenomenological assignments of the scalar masses are discussed in
the Particle Data Group (PDG) review on scalar mesons of Ref.~\cite{PDG}.}
\vspace*{-0.25cm}
\end{table}
\begin{equation}
2\sqrt{2} f_K=\sqrt{2} x+{{2}}y, \quad\quad f_\pi=x. \label{eq:pcac}
\end{equation}
Equations (\ref{eq:ex}), (\ref{eq:ey}), (\ref{eq:pcac}) and those of
Table~\ref{Tab:masses} connect at tree--level the eight
parameters of the Lagrangian 
($x$, $y$, $\mu_0$, $f_1$, $f_2$, $g$, $\epsilon_x$, $\epsilon_y$) 
and the physical characteristics of the meson sector. 
$x$ and $y$ belong to the coupling parameters of the
shifted Lagrangian.
The pseudoscalar masses and decay constants are better known than the
corresponding quantities of the scalar sector, therefore the pseudoscalar
sector is preferred over the scalars for fixing the parameters. The $x$,
$y$ condensates are simply obtained from (\ref{eq:pcac}). The couplings
$f_2$, $g$ and the combination ${M}^2:=-\mu_0^2+4f_1(x^2+y^2)$  
can be determined by the knowledge of three pseudoscalar masses. The pion
and kaon masses obviously should be selected since our purpose is to
study the
effect of their variation on the thermodynamics. For the third physical
quantity, the trace of the mass matrix in the $\eta$-sector is
chosen, which will be denoted below by $M_\eta^2$.

This set of relations has the following explicit solution:
{\allowdisplaybreaks
\begin{eqnarray}
x&=&f_\pi \, , \label{x} \\
y&=&\left(2f_K-f_\pi\right)/\sqrt{2} \, , \label{y}  \\
f_2 &=&
\frac{(6f_K-3f_\pi)m_K^2-(2f_K+f_\pi)m_\pi^2-2(f_K-f_\pi){M}_\eta^2}
{4(f_K-f_\pi)(8f_K^2-8f_K f_\pi+3f_\pi^2)} \, ,\label{f2} \\
g&=& \frac{2f_K
  m_K^2+2(f_K-f_\pi)m_\pi^2-(2f_K-f_\pi){M}_\eta^2}{\sqrt{2}
(8f_K^2-8f_K f_\pi+3f_\pi^2)} \label{g} \, , \\ 
M^2&=&\frac{1}{2}M_\eta^2+\frac{g}{\sqrt{2}}(2f_K-f_\pi)-
2 f_2 (f_\pi (f_\pi-2 f_K)+2 f_K^2).
\label{Mbar} 
\end{eqnarray}
}\noindent
The above relations contain informations only from the pseudoscalar sector
and were previously used in Ref. \cite{Lenaghan00}.  

$\epsilon_x$ and $\epsilon_y$ are
determined when using the above expressions in (\ref{eq:ex}) and
(\ref{eq:ey}). But it is simpler to combine these equations in the 
Gell-Mann--Oakes--Renner (GMOR) relations and use the following equations
instead of the equations of state:
\begin{equation}
\epsilon_x=m_\pi^2 x,  \quad \quad
\epsilon_y=\frac{\sqrt{2}}{2}
(m_K^2-m_\pi^2)x+m_K^2y \label{ae1}.
\end{equation}
These tree--level Ward-identities guarantee the Goldstone theorem at zero
temperature. When $m_\pi^2=0$, the external field $\epsilon_x$ is zero and
$\epsilon_y$ generates the nonzero value of $m_K$. 
The approach of taking into account the variation of the pion and kaon mass
only by changing the external fields (cf. (\ref{ae1}) ) was extensively followed in
the recent literature, see e.g. \cite{roder03,Lenaghan00,Meyer-Ortmanns}.

The combination ${M}^2$ of $f_1$ and $\mu_0^2$ is split up only in the
expression of the admixed scalars, therefore the use of one characteristics
of the mixed scalar spectra is unavoidable \cite{Tornqvist99}. 
Nothing is known about the dependence of the $\sigma$ mass on $m_\pi$
and $m_K$. We have applied the method described in detail in Section
\ref{sec:termo} also to the case when the mass of the $\sigma$ mass
was fixed to a single value in the entire $(m_\pi - m_K)$--plane. This
scheme results in a phase diagram which is not compatible with the
universal arguments on the nature of the phase transition in the
chiral limit, at least for smaller sigma mass values, preferred
nowadays \cite{low_sigma}. Therefore some more flexible relation
should be tested which allows the variation of the $\sigma$ mass with
the pseudoscalar masses.  We explored the consequences of assuming two
different relations for the mass matrix of the scalars in the
($m_\pi-m_K$)--plane: {\allowdisplaybreaks
\begin{itemize}
\item[A1.]
A first alternative is to assume that the mixing in the scalar $x-y$ sector 
is absent ($m^2_{\sigma_{xy}}=0$), which along the $m_\pi=m_K$ line 
is the consequence of the $U(3)\times U(3)$ Gell-Mann--Okubo (GMO) relation.
\item[A2.]
The $SU(3)~\times~SU(3)$ GMO mass formula for the scalars (\ref{sgmo}) is
fulfilled in the physical point with an accuracy of about $-1.7\%$,
supposing $m_\sigma=600$ MeV. A second alternative is to require it to be
fulfilled with the same accuracy for arbitrary $m_\pi, m_K$.
\end{itemize}}\noindent
Both assumptions involve a certain arbitrariness. The phase diagram
was mapped out using both alternatives, and the resulting deviations 
give some feeling of the effects of our ignorance concerning the
scalar sector.

We give here the expression of $f_1$ and $\mu_0^2$ for the alternative ,A1'
applied to the scalar sector:
\be
f_1^{(A1)}=-\frac{g}{4 y},\qquad
{\mu_0^2}^{\,(A1)}=4f_1^{(A1)}(x^2+y^2)- {M^2},
\label{f1_mu_A1} 
\ee
where the superscript ,A1' refers to the nonmixing of the $x-y$ scalars.
We can see in the equation above that in alternative ,A1' the coupling
$f_1$ is directly proportional to the strength of the $U_A(1)$ breaking 
determinant term in the Lagrangian (\ref{Lagrangian}).
The implementation of the assumption ,A2' is more complicated, hence it is
detailed in Appendix \ref{App:A2}.

The logics of the procedure sketched above can be summarized as follows: 

\begin{tabular}{p{5.5cm} p{1.5cm} p{0.1cm} p{1.5cm} p{0.2cm} p{1.5cm}}
 & {\bf input:} & & {\bf output:} & & {\bf prediction:} \\
\end{tabular}
\begin{equation*}
\begin{aligned}
  \left.
  \begin{aligned}
     \left.
     \begin{aligned}
      f_\pi \, \,  \\ f_K  \, \,
     \end{aligned}
      \right\} & \Longrightarrow &  \begin{aligned}
                                    x\quad \, \\
                                    y\quad \,
                                    \end{aligned}
                                    \\
     \left.
     \begin{aligned}
      m_\pi  \,  \\ m_K \,  \\ {M_\eta^2}   
      \end{aligned}
      \right\} & \Longrightarrow &   \begin{aligned}
                                     g\quad \, \\ 
                                     f_2\quad \\ 
                                     {M}^2 \, \,\,
                                     \end{aligned}
      \end{aligned} \right\}                             \Longrightarrow  
 \begin{aligned}
 m_\eta \quad\quad \quad \quad\\
m_{\eta^{'}}\,\,\,\, \quad \quad \quad \\
\theta_\eta  \quad  \quad \quad \quad \\
m_{a_0}\,\,\, \quad \quad \quad \\
m_{\kappa}\,\, \,\,\quad \quad \quad
\end{aligned} \\
 \begin{aligned}
\left. 
\begin{aligned}
  \textnormal{A1 \& ${M}^2$}\,  \\ \textnormal{A2 \& ${M}^2$}\, 
\end{aligned}
\right\}
 & \Longrightarrow & \left.
\begin{aligned}
\mu_0^2\,\,\,\,\, \\ f_1 \, \,\,\,\,
\end{aligned}
\right\}   \Longrightarrow  
\,m_\sigma,\,  m_{f_0} ,\, \theta_\sigma \, \\
\left. 
\begin{aligned}
E_x=0\,\,\, \\ E_y=0 \,\,\,
\end{aligned}
\right\} 
& \Longrightarrow &
\begin{aligned}
\epsilon_x \quad\quad \quad\quad\quad \quad\,\,\, \quad\quad \quad\\ 
\epsilon_y \quad \quad \quad \quad\quad \quad\,\,\,\quad\quad \quad
\end{aligned}
\end{aligned}
\end{aligned}
\end{equation*}
where $m_\sigma$, $m_{f_0}$
are the mass eigenvalues of the admixed scalars and $\theta_\sigma$ is their
mixing angle.

The dependence of the parameters on $m_\pi$ and $m_K$ in
(\ref{x})\,-(\ref{ae1}) is not only explicit because one learns from the 
Chiral Perturbation Theory (ChPT) that all physical quantities 
($f_\pi$, $f_K$, ${M}_\eta^2$) featuring in this expressions also depend on 
$m_\pi, m_K$. Consequently, for the parametrization of the effective 
sigma model for arbitrary $m_\pi$, $m_K$ one should use the correct 
$f_\pi(m_\pi,m_K)$, $f_K(m_\pi,m_K)$, ${M}_\eta^2(m_\pi,m_K)$ functions. 
In the next section we will construct these functions relying on results of 
the three-flavor ChPT.


\section{Dependence of the couplings on $m_\pi$ and $m_K$}

The fundamental problem of effective models in exploring the phase
diagram of QCD in the $(m_\pi - m_K)$--plane is the determination of
the variation of the effective couplings when moving in the plane. The
values determined in the physical point serve only as reference
points, for a systematic exploration some reliable external reference
is needed.

The situation is somewhat analogous (but reciprocal!) to lattice QCD,
where simulations are performed in a range of quark masses leading to
much heavier pseudoscalars than in nature and some guidance is needed
to arrive to the physical point.  Chiral perturbation theory (ChPT) is
used in this extrapolation \cite{chiral_extrapolation}. Very recently
it was applied in \cite{Meisner}, for analyzing the pion 
mass dependence of the baryon masses of MILC collaboration. This suggests
to us the idea to make use of ChPT results for deriving the
parametrization of the linear sigma model away from the physical
point. The issue of the compatibility of $L\sigma M$ and ChPT is not
entirely settled. Recently the two models were compared in
\cite{Bramon} in the light of the latest experimental data.  The
information available on the scalar sector, which improved
considerably in the past few years, was used to fix some of the low
energy constants of the ChPT with a more satisfactory result than
thought possible previously.  In this paper we fix the low energy
constants within the ChPT, and adjust its renormalization scale, in
order to match the pseudoscalar masses of the nonlinear sigma model
with the tree--level spectra of $L\sigma M$ over an extended range of
the $(m_\pi - m_K)$-plane. The calculated low energy constants of ChPT
fall in the range commonly used in the literature.

The essence of our approach can be understood by restricting our attention
first to the functions $f_\pi(m_\pi,m_K)$ and $f_K(m_\pi,m_K)$ (the
$\eta -\eta'$ mixing will be discussed afterwards). For
this purpose, it is sufficient to choose the framework of 
$SU(3)~\times~SU(3)$ ChPT \cite{Gasser85}. There  8 parameters
 ($f, A, q, M_0, L_4, L_5, L_6, L_8$) were introduced,
which determine $m_\pi^2,
m_K^2, f_\pi, f_K$ with ${\cal O}(1/f^2)$ accuracy:
{\allowdisplaybreaks
\bea
\label{pi-mass}
m_\pi^2&=&2A\left[1+\frac{1}{f^2}\left(\mu_\pi-\frac{1}{3}\mu_\eta
+16A(2L_8-L_5)+16A(2+q)(2L_6-L_4)\right)\right],\\
\label{k-mass}
m_K^2&=&A(1+q)\left[1+\frac{1}{f^2}\left(\frac{2}{3}\mu_\eta
+8A(1+q)(2L_8-L_5)+16A(2+q)(2L_6-L_4)\right)\right],\\
\label{pi-decay}
f_\pi&=&f\left[1+\frac{1}{f^2}\left(-2\mu_\pi-\mu_K+8AL_5+8A(2+q)L_4\right)
\right],\\
f_K&=&f\left[1+\frac{1}{f^2}\left(-\frac{3}{4}(\mu_\pi+\mu_\eta+2\mu_K)+
4A(1+q)L_5+8A(2+q)L_4\right)\right],
\label{k-decay}
\eea
}\noindent
where
$\mu_{PS}=m_{PS}^2\ln(m_{PS}^2/M_0^2)/(32\pi^2)$
are the so--called chiral logarithms at scale $M_0$, in which $m_{PS}^2$ is
substituted by the leading order expression for the squared mass of
the corresponding member of the pseudoscalar octet.
To this order one has in agreement with the Gell-Mann--Okubo formula
$m_\eta^2=2A(1+2q)/3$.
It is worth to emphasize that $L_i$ do not vary 
with the pseudoscalar masses.

The parameters $A$ and $q$ are related directly to the quark masses
(cf. \cite{Gasser85}) through
$A=B(m_u+m_d)/2$ and $q=2m_s/(m_u+m_d)$
where $B$ is determined by the condensate $\lag\bar u u\rag$ in the chiral
limit.  They can be expressed readily 
through the pseudoscalar masses and the chiral
constants $L_i$  by `inverting'' Eqs. (\ref{pi-mass}) and (\ref{k-mass})
to ${\cal  O}(1/f^2)$ accuracy:
\bea
\label{Acont}
A&=&\frac{m_\pi^2}{2}\left[1-\frac{1}{f^2}\left(\mu_\pi-\frac{1}{3}
\mu_\eta+8m_\pi^2(2L_8-L_5)+8(2m_K^2+m_\pi^2)(2L_6-L_4)\right)\right],\\
1+q&=&\frac{2m_K^2}{m_\pi^2}\left[1-\frac{1}{f^2}\left(\mu_\eta-
\mu_\pi+8(m_\pi^2-m_K^2)(2L_8-L_5)\right)\right].
\label{xcont}
\eea
It is sufficient to use the leading order relations of the two equations
above to extract from  Eqs. (\ref{pi-decay}) and (\ref{k-decay})
 the following $m_\pi, m_K$-dependence for the pseudoscalar 
decay constants:
\bea
f_\pi&=&f\left[1-\frac{1}{f^2}(2\mu_\pi+\mu_K-4m_\pi^2(L_4+L_5)-
8m_K^2L_4)\right],
\label{fpi-extrap}\\
f_K&=&f\left[1-\frac{1}{f^2}\left(\frac{3}{4}(\mu_\pi+\mu_\eta+2\mu_K)-
4m_\pi^2L_4-4m_K^2(L_5+2L_4)\right)\right].
\label{fk-extrap}
\eea
Using as input $f_\pi=93$ MeV, $f_K=113$ MeV, $m_\pi=138$ MeV, 
$m_K=495.6$ MeV, and $m_\eta=547.8$  MeV, and  choosing
\be
M_0=4\pi f_\pi\approx 1168\ \textrm{MeV},\qquad f=88\ \textrm {MeV},
\ee
one finds in the physical point the following
values for the relevant chiral constants:
\be
L_4=-0.7044\times 10^{-3},\qquad L_5=0.3708\times 10^{-3},
\label{l4l5}
\ee
which completes the continuation formulas for the decay constants
(\ref{fpi-extrap}) and (\ref{fk-extrap}). 

\begin{figure}[htpb]
\centering{
\includegraphics[width=0.45\textwidth, keepaspectratio]{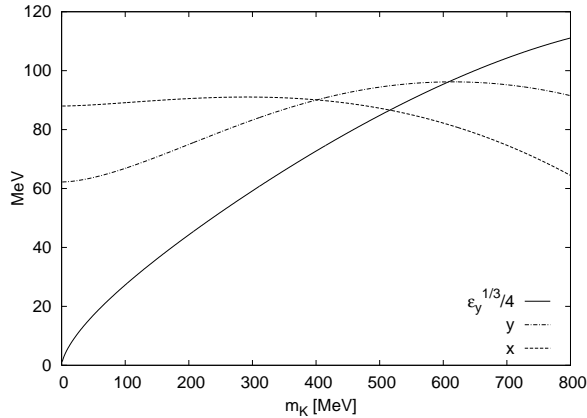}}
\caption{The tree--level kaon mass dependence of the $T=0$ condensates $x$ and 
$y$ and the external field $\epsilon_y$ for $m_\pi=0$ ($\epsilon_x=0$).} 
\label{fig:x_y}
\end{figure}

These formulas enable us to predict the mass variation of the chiral
condensates with help of Eqs. (\ref{x}) and (\ref{y}),
and also the external fields $\epsilon_x,\epsilon_y$ from (\ref{ae1}). 
The $m_K$ dependence of $x,y$, and $\epsilon_y$ is displayed for $m_\pi=0$ in 
Fig.~\ref{fig:x_y}. 
We remark that the only attempt, we are aware of, to take into account 
the nontrivial mass dependence of $f_\pi(m_\pi)$ in a thermal analysis,
was based on fitting and extrapolating the mass dependence measured on lattice
\cite{Lenaghan00}.

The chiral constants $L_6$ and $L_8$ are controlled by the values of $A$ and
$q$, respectively, taken in the physical point. Especially simple is the 
relation of $L_8$ to the ratio $q$ of the strange to average nonstrange 
quark mass. We take the value $q=24.8$, which is close 
to the lattice determination and compatible with the range indicated by the
\mbox{PDG listing \cite{PDG}: $20 \lesssim q\lesssim 34$.} For $A$ we choose
its leading order ChPT value in the physical point: $A=A^{(0)}$.
Then using in the ${\cal O}(1/f^2)$ accurate expressions of $A$ and
$q$  the phenomenological values of
$m_\pi^2, m_K^2$ with the Gell-Mann--Okubo formula for $m_\eta^2$ one obtains
\be
L_6=-0.3915\times 10^{-3},\qquad L_8=0.511\times 10^{-3}.
\ee
The values of the chiral constants $L_i$, together with $M_0$ and $f$
can be used further for the continuation of $A$ and $q$ from
the physical point to an arbitrary point of the ($m_\pi - m_K$)--plane.

The complete $m_\pi, m_K$-dependence of the couplings $f_2, g, M^2$ given in
Eqs.~(\ref{f2}), (\ref{g}), (\ref{Mbar}) requires also the knowledge of $
M^2_\eta(m_\pi, m_K)$, that is the mass dependence in the ($\eta_0,
\eta_8$)-sector, for which the application of $U(3)\times U(3)$ ChPT is
needed. The steps are quite analogous to what was described above, but the
mass mixing makes it somewhat complicated. Since these formulae can be
found dispersed in several papers we collect here the relevant
formulae in more detail.

In this sector, the ${\cal O}(1/f^2)$ ChPT results in a Lagrangian of the
following form \cite{Herrera98,Herrera97,Borasoy01,Beisert01}:
\be
L_{08}=\frac{1}{2}A_{ij}\partial_\mu\eta_i\partial^\mu\eta_j-
\frac{1}{2}D_{ij}\eta_i\eta_j,\qquad i=0,8,
\ee
where the elements of the real symmetric matrices $A$ and $D$ 
\be
A_{ij}=\delta_{ij}+a_{ij},\qquad D_{ij}=D_{ij}^{(0)}+d_{ij} .
\ee
can be read off the papers \cite{Borasoy01,Beisert01} and are compiled in
Appendix B for the reader's convenience. The matrices $a_{ij}, d_{ij}$
represent ${\cal O}(1/f^2)$ corrections to the zeroth order quantities.

This Lagrangian is diagonalized in two steps. First one redefines the
two-component vector $\eta_i$ as $\tilde\eta_i:=A^{1/2}_{ij}\eta_j$
which is followed by an appropriate  rotation $R(\theta_\eta)\tilde\eta$:
\bea
&&\begin{pmatrix}  \eta\\ \eta'  
\end{pmatrix}=R(\theta_\eta)\left(1+\frac{1}{2}a\right)
\begin{pmatrix}  \eta_8\\ \eta_0  \end{pmatrix}, \nonumber\\
&&\begin{pmatrix}  m^2_\eta & 0 \\ 0 &  m^2_{\eta'}  \end{pmatrix}=
R(\theta_\eta)
\begin{pmatrix}  m^2_{\eta_{88}}   & m^2_{\eta_{08}}\\ m^2_{\eta_{08}}
  & m^2_{\eta_{00}}
\end{pmatrix}R^{-1}(\theta_\eta)
=R(\theta_\eta)\left(1-\frac{1}{2}a\right)D
\left(1-\frac{1}{2}a\right)R^{-1}(\theta_\eta). \label{forg}
\label{Ratrans}
\eea 

Choosing $\theta_\eta=-20^\circ$ and the experimental information on $m_\eta,
m_{\eta'}$ one finds in the physical point the values of
$m_{\eta_{00}}^2, m_{\eta_{08}}^2, m_{\eta_{88}}^2$, which represent
three relations restricting four chiral constants $L_7, v_0^{(2)}, v_2^{(2)},
v_3^{(1)}$ appearing in the respective ChPT expressions for their
masses. We choose the large $N_c$ relation $v_0^{(2)}=-29.3f^2$
\cite{Herrera98} in order to have as many unknown chiral
constants as relations among them. This constant represents the contribution
of the $U_A(1)$ anomaly to the $\eta$ mass, dominantly determined by the
topological features of the gluon configurations. It should be rather
insensitive to the variation of the quark masses.
From the expressions of the mass matrix elements 
listed in (\ref{etamass})-(\ref{etamass3}) one finds for the
chiral constants:
\be
L_7=-0.2272\times 10^{-3},\qquad v_3^{(1)}=0.095, \qquad
v_2^{(2)}=-0.1382.
\ee
In the parametrization of $L\sigma M$
the sum of equations (\ref{Eq:m_eta88}) and (\ref{Eq:m_eta00}) is used: 
\begin{eqnarray}
{M_\eta^2}&=&2m_K^2-3v_0^{(2)}+2(2m_K^2+m_\pi^2)
(3v_2^{(2)} - v_3^{(1)}) +\frac{1}{f^2}\left[8v_0^{(2)}
(2m_K^2+m_\pi^2)(L_5+3L_4)+m_\pi^2(\mu_\eta-3\mu_\pi )
-4m_K^2\mu_\eta\right.\nonumber \\
&+&\left.\frac{16}{3}(6L_8-3L_5+8L_7)(m_\pi^2-m_K^2)^2+
\frac{32}{3}L_6(m_\pi^4-2m_K^4+m_K^2m_\pi^2)+
\frac{16}{3}L_7(m_\pi^2+2m_K^2)^2 \right] \label{Meta-extrap} .
\end{eqnarray}

\begin{figure}[htbp]
\begin{minipage}[t]{0.49\textwidth}
\centering {\includegraphics[width=0.99\textwidth, keepaspectratio]
{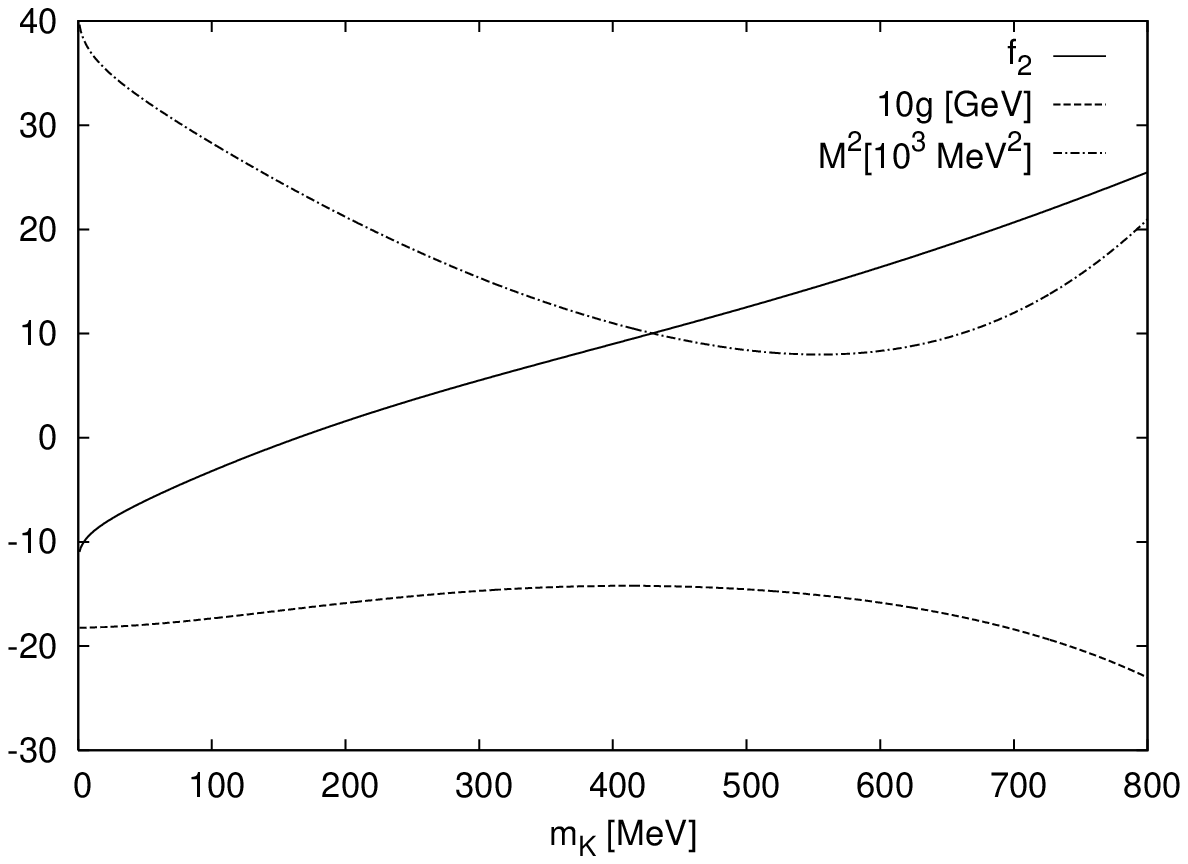}}
\caption{The tree--level kaon mass dependence of the parameters of L$\sigma$M
determined solely from the pseudoscalar sector: 
$f_2, g,$ and $M^2$ for $m_\pi=0$.} 
\label{fig:f2_g_M}
\end{minipage}%
\hfill
\begin{minipage}[t]{0.49\textwidth}
\setlength{\abovecaptionskip}{0pt}
\centering {\includegraphics[width=0.99\textwidth, 
keepaspectratio]{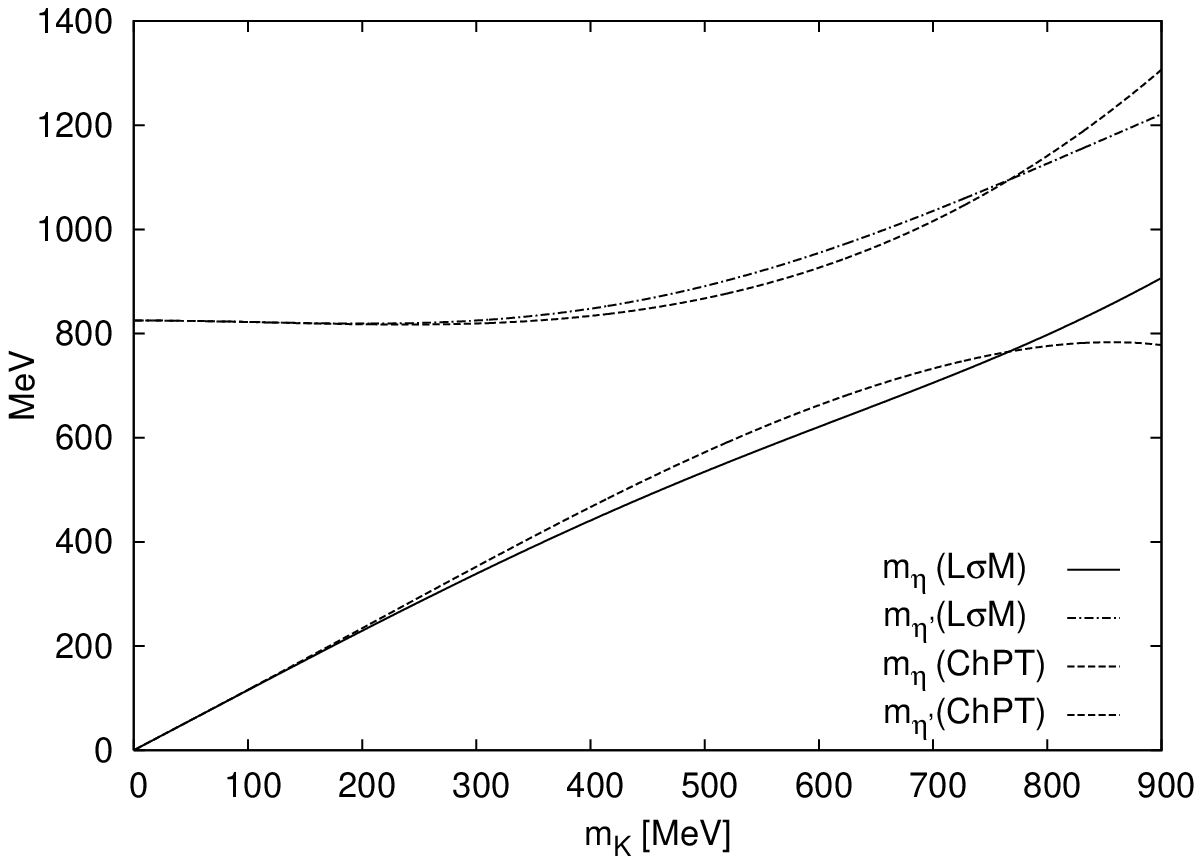}  }
\caption{The tree--level kaon mass dependence of $m_\eta$ and $m_{\eta^{'}}$ 
for $m_\pi=0$. The labels refer to the results of ChPT and the predictions
 of linear sigma model (L$\sigma$M), respectively.}
\label{fig:m_eta}
\end{minipage}%
\end{figure}

It can be checked that our results (\ref{fpi-extrap}),
(\ref{fk-extrap}) and (\ref{Meta-extrap}) are the same as in
\cite{Herrera98}, when the $\mu_{PS}$'s and $L_4$, $L_6$, $L_7$,
$v_2^{(2)}$ are set equal to zero (corresponding to the large $N_c$
limit). The chiral logarithm $\mu_\eta$ contains the $\eta$ mass at
leading order: $(m_\eta^{(0)})^2=(4m_K^2-m_\pi^2)/3$, therefore the
functions $M_\eta^2(m_\pi,m_K)$ and $f_\pi(m_\pi,m_K)$,
$f_K(m_\pi,m_K)$ are only applicable when $4m_K^2 > m_\pi^2$.  In
addition we can rely on our ``classical'' approximation if the masses
are lower than the chiral scale $M_0$.  Eq.~(\ref{Meta-extrap})
together with (\ref{fpi-extrap}) and (\ref{fk-extrap}) allows the
computation of the couplings $f_2, g, M^2$ in the pseudoscalar 
mass--plane. Their variation is illustrated in Fig.~\ref{fig:f2_g_M} for
$m_\pi=0$.  The theoretical quality of this parametrization is
illustrated here by comparing $m_\eta (m_K, m_\pi=0)$ and
$m_{\eta'}(m_K, m_\pi=0)$ as computed from the tree--level expressions
of the linear sigma model with the results for the same quantities
directly obtained from ChPT. Fig.~\ref{fig:m_eta} demonstrates that up
to $m_K= 800$ MeV the agreement is almost perfect.

\begin{figure}[htbp]
\vspace*{5pt}
\begin{minipage}[t]{0.49\textwidth}
\centering {\includegraphics[width=0.99\textwidth, 
keepaspectratio]{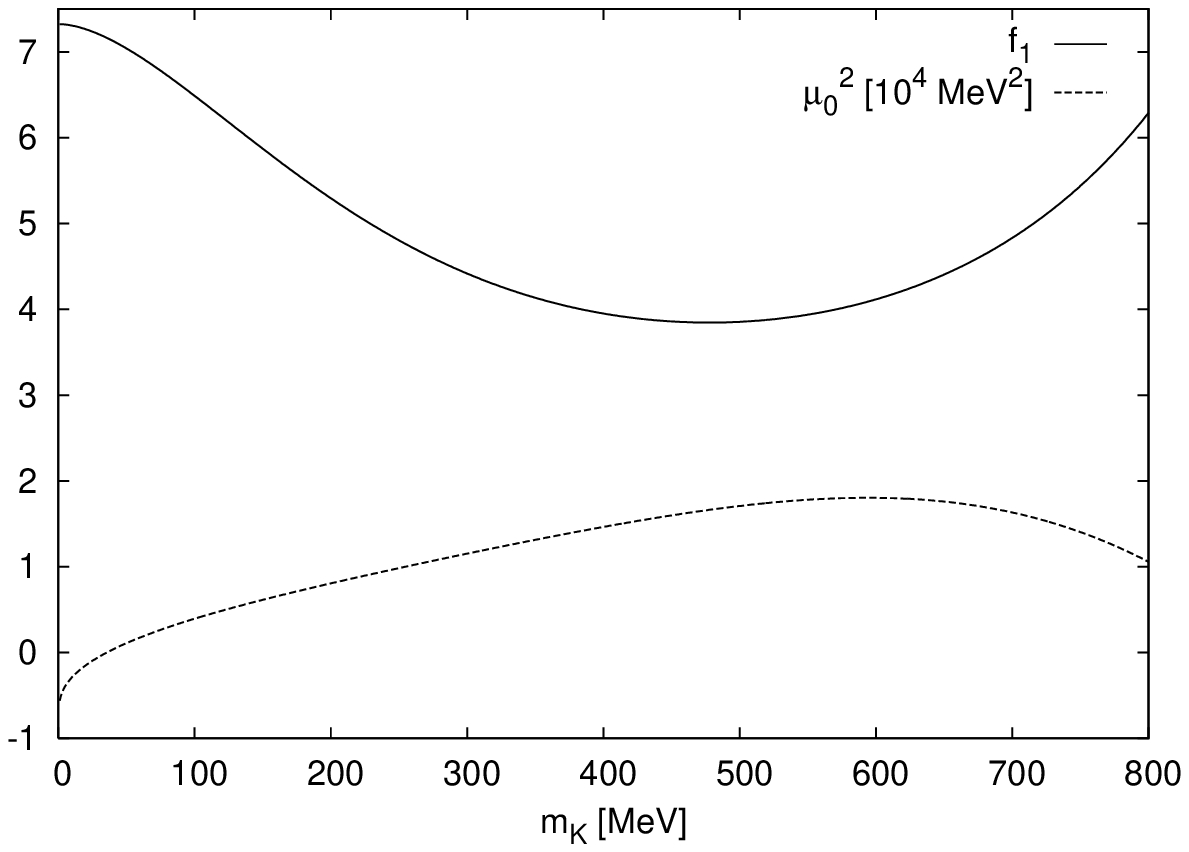}}
\caption{The tree--level kaon mass dependence of $f_1$ and $\mu_0^2$ 
with alternative ,A1' for the scalar sector, when $m_\pi=0$.  } 
\label{fig:mu_f_1}
\end{minipage}%
\hfill
\begin{minipage}[t]{0.49\textwidth}
\centering {\includegraphics[width=0.99\textwidth, 
keepaspectratio]{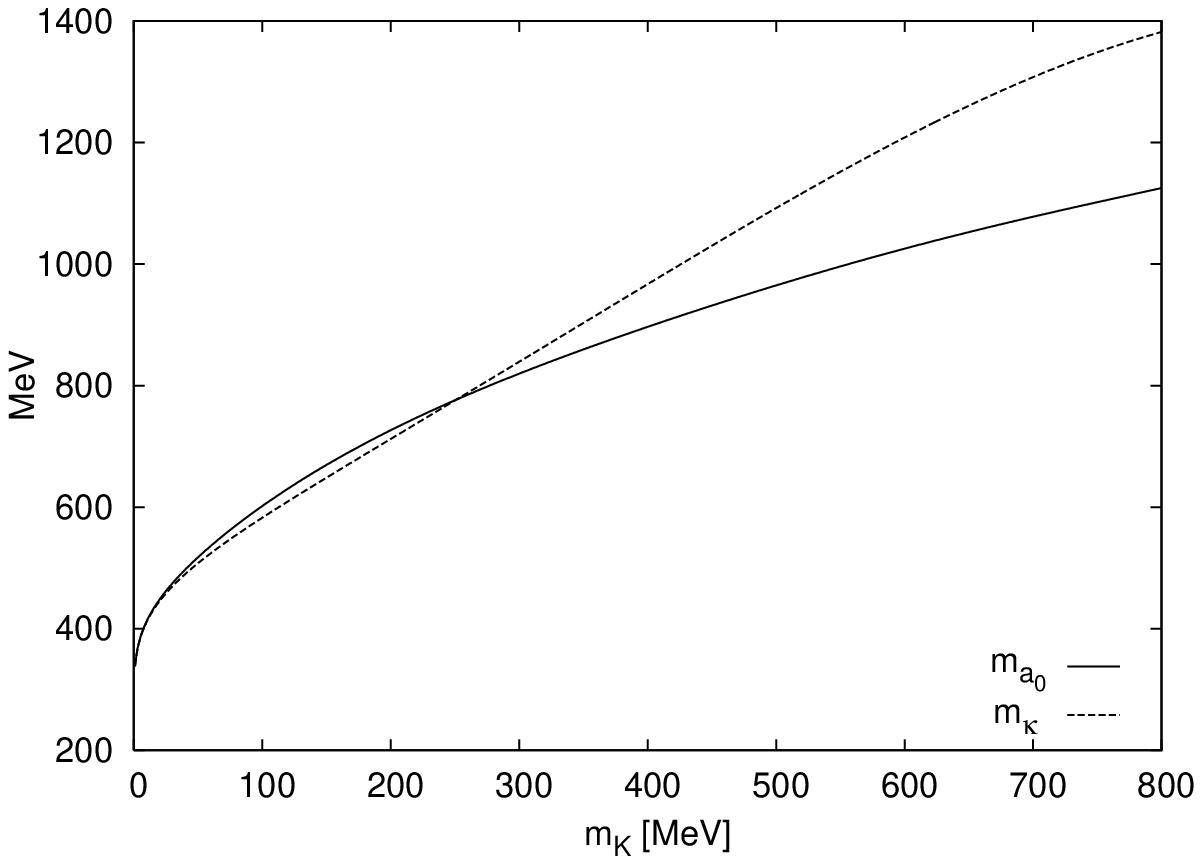}}
\caption{The tree--level kaon mass dependence predicted for scalar meson
masses $a_0$ and $\kappa$ in the linear sigma model, when $m_\pi=0$.}
\label{fig:m_a0kappa}
\end{minipage}%
\end{figure}

The splitting of $M^2$ into $f_1$ and $\mu_0^2$ is realized in the
mixing scalar sector, therefore it does not require any further
consideration of ChPT. Their curves are shown for
alternative ,A1' in Fig.~\ref{fig:mu_f_1},
while the predicted masses of $a_0$ and $\kappa$ are given in 
Fig.~\ref{fig:m_a0kappa}. 
It turns out that the alternative parametrization ,A2'
leads to divergences in $f_1$ and $\mu_0^2$ for $m_K-m_\pi\lessapprox
200$ MeV. Therefore one cannot use it for the exploration of the
whole ($m_\pi - m_K$)--plane.

A final remark concerns the sensitivity of the $T=0$ mass
spectra relative to the chiral constants ($L_i,v_i^{(j)}$). The values of the 
constants change considerably if, for instance, the large $N_c$ limiting
formulas of ChPT are used. This change results in a rather large variation in
the numerical values of $f_2,g,M^2$. However, the predicted masses of
$\eta, \eta'$ and the scalar sector remain almost unchanged.

With this novel $m_\pi - m_K$-sensitive parametrization of the linear
sigma model we are going to discuss the nature of the temperature
driven chiral symmetry restoration in the following sections.

\section{Quasiparticle thermodynamics of the $SU(3)~\times~SU(3)$ model
\label{sec:termo}}

The aim of this section is to derive the equations of state (EoS)
which determine the variation of the order parameters $x$ and $y$ with
the temperature, including the existence of multiple solutions in
certain temperature ranges. Results of the numerical analysis of EoS
determining the nature of the transition in function of the masses
$m_\pi, m_K$ will be mapped out in the next section.

The renormalized EoS will be determined in the framework of Optimized
Perturbation Theory of Chiku and Hatsuda \cite{chiku} which starts by
reshuffling the mass term of the Lagrangian density by introducing a
temperature dependent effective mass parameter:
\be
L_{mass}=-\frac{1}{2}M^2(T)\tr M^\dag M
+\frac{1}{2}(\mu_0^2+M^2(T))\tr M^\dag M.
\label{resum}
\ee
The first term on the right hand side is used in the thermal propagators 
of the different mesons. The second term in (\ref{resum}) represents the 
effective mass counterterm which is taken into account in higher orders 
of the perturbative calculations.

The tree--level mass of $\pi$ involves now the thermal mass parameter:
\be
m_\pi^2=M^2(T)+2(2f_1+f_2)x^2+4f_1y^2+2gy,
\label{resum-pion}
\ee
and all other meson masses to be used in the tadpole integrals below
agree with the formulas appearing in Table~\ref{Tab:masses} with the
replacement $-\mu^2_0\rightarrow M^2(T)$.  If all quantum corrections
are condensed into $M^2(T)$, then the tree--level masses of other
mesons are expressible through the mass of the pion.  One might expect
that the pion has the lowest mass and therefore for $M^2(T) > 0$ these
squared masses are all positive, which is not the case when
$-\mu_0^2<0$ figures in the propagators.  We define a physical region
of $x$ and $y$ where all tree--level mass squares are positive, and
thus the one-loop contribution of the meson fluctuations to EoS is
real. This region is most severely restricted by the masses of $f_0$ and
$\sigma$, which strongly decrease near the phase transition. We will
look for the solution of the EoS's in the physical region.

For the determination of the thermal mass we use the Schwinger-Dyson
equation for the inverse pion propagator at zero external momentum.
At one-loop it  receives the contribution $\Pi(M(T),p=0)$, which is the
self-energy function of the pion at zero external momentum, plus the
counterterm contribution $-\mu_0^2-M^2(T)$. We apply the principle of 
minimal sensitivity (PMS) \cite{chiku}, that is we require that the pion 
mass be given by its tree--level expression:
\be
\label{Eq:M_T_gap}
\Pi_\pi(M(T),p=0)-\mu_0^2-M^2(T)=0.
\ee
$\Pi(M(T),p)$ itself is a linear combination of the tadpole and bubble
diagrams (the latter not included in the treatment of
\cite{Lenaghan00}), with coefficients derived with help of the 4-point
and 3-point couplings among mass eigenvalue fields. This step requires
diagonalization in the $(x,y)$ sector cf. Appendix \ref{App:t_coeff}. 
The bubble contribution
$B(m_1,m_2,T,p=0)$ at zero external momentum $p$ can be expressed
through tadpole integrals $I(m_i,T)$ as
\be
\label{Eq:split}
B(m_1,m_2,T,p=0)=\frac{I(m_1,T)-I(m_2,T)}{m_1^2-m_2^2},
\ee
therefore the self-energy is easiest to represent in form of a linear
combination of tadpole integrals, which gives when substituted into 
Eq.~(\ref{Eq:M_T_gap}):
\begin{equation}
0=-M^2(T)-\mu_0^2+
\sum^{\alpha=\sigma,\,\pi}_{i=\pi,\,K,\,\eta,\eta^{'}} c^\pi_{\alpha_i
} I(m_{\alpha_i}(T),T) \, . \label{gap}
\end{equation}
Here $c^\pi_{\alpha_i}$ are the weights of the tadpole contributions
evaluated with different mass eigenstate mesons $\alpha_i=\sigma_i, \pi_i$.
The integrals over the corresponding propagators are
evaluated with effective tree--level masses where $M^2(T)$ replaces
$-\mu_0^2$. In this way (\ref{gap})
is actually a gap equation which determines the thermal mass
parameter, $M^2(T)$. With help of Eq. (\ref{resum-pion}) this
equation can be also understood as a gap equation for the pion mass
(the pion mass is present also in the expressions of
$I(m_{\alpha_i},T)$ through $m_{\alpha_i}$!):
\be
m_\pi^2=-\mu_0^2+2(2f_1+f_2)x^2+4f_1y^2+2gy+\sum^{\alpha=
\sigma,\,\pi}_{i=\pi,\,K,\,\eta,\eta^{'}} c^\pi_{\alpha_i
} I(m_{\alpha_i}(T),T) \, .
\label{pigap}
\ee
Since this equation depends also on the order parameters $x,y$ we
have to solve in addition to the gap equation the two equations of state: 
\begin{eqnarray}
\label{eq:EoS_x}
&-\epsilon_x-\mu_0^2 x+2gxy+4f_1xy^2+2(2f_1+f_2)x^3+
\sum^{\alpha=\sigma ,\pi}_{i=\pi,\,K,\,\eta,\eta^{'}}  J_i t^x_{\alpha_i
} I(m_{\alpha_i}(T),T)&=0 \label{xeq}\,,\\
\label{eq:EoS_y}
&-\epsilon_y-\mu_0^2 y+gx^2+4f_1x^2y+4(f_1+f_2)y^3+
\sum^{\alpha=\sigma ,\pi}_{i=\pi,\,K,\,\eta,\eta^{'}}  J_i t^y_{\alpha_i
} I(m_{\alpha_i}(T),T)&=0\,, \label{egy2}
\end{eqnarray}
with $t^x_{\alpha_i}$ and $t^y_{\alpha_i}$ giving the corresponding
weights, listed in Appendix C. $J_i$ is the isospin multiplicity factor: 
$J_\pi=3$, $J_K=4$, and  $J_{\eta,\eta'}=1$.   

Equations (\ref{gap}), (\ref{xeq}) and (\ref{egy2}) represent a polynomial in
$x,y$ with divergent coefficients due to the divergences of the
tadpole integral $I$. When compared to the expressions of the tree--level 
pion mass in Table~\ref{Tab:masses} and the tree--level EoS (\ref{eq:ex}),
(\ref{eq:ey}) one can uniquely absorb divergences into the couplings
$-\mu_0^2, f_1, f_2, g$. This step requires divergent counterterms as follows:
{\allowdisplaybreaks
\begin{eqnarray*}
\delta \mu_0^2 &= & \frac{   (5f_1+3f_2)\Lambda^2}{\pi^2}-  
\frac{(5f_1+3f_2)M^2(T)-g^2}{\pi^2}\ln\frac{\Lambda^2}{l^2}, \label{mur} \\
\delta g&=&\frac{3g(f_1-f_2)}{2\pi^2}  \ln\frac{\Lambda^2}{l^2},  \\
\delta f_1&=&\frac{13f_1^2+12f_1 f_2+3f_2^2}{2\pi^2}
 \ln\frac{\Lambda^2}{l^2}, \\
\delta f_2&=&\frac{3f_1 f_2+3f_2^2}{\pi^2} \ln\frac{\Lambda^2}{l^2},  
\label{fkren}
\end{eqnarray*}
}\noindent
where $\Lambda$ is the regularization cut-off and $l$ is the
renormalization scale.  At $T=0$ with the replacement
$M^2(T)\rightarrow~-~\mu_0^2$ these expressions agree with the known
coupling renormalizations \cite{Haymaker73}. $T$-dependence appears
only in the mass renormalization, through $M(T)$. Since they are
proportional to higher powers of the couplings, this apparent
environment dependence of the counterterm will be canceled by
higher--loop contributions (see for instance, \cite{jako05}).  At the
end of the renormalization we arrive at the same equations, just one
has to replace $\mu_0^2, f_1,f_2, g, I$ by their renormalized
expressions (separate notation will be introduced below only for
$I\rightarrow I_R$).

The coefficients $c^\pi_{\alpha_i}$ look at first sight horribly
complicated since not only specific three-point couplings (see
Appendix C) but also weighted factors proportional to
$(m_{\sigma_i}^2-m_{\pi_j}^2)^{-1}$ do contribute,
cf. Eq.~(\ref{Eq:split}). However, a wonderful simplification occurs
when working through this complicated expression, one finds
$c^\pi_{\alpha_i}=J_i t^x_{\alpha_i}/x$.  Then comparing the gap
equation (\ref{pigap}) to the EoS for the order parameter $x$, one
recognizes the relation
\be
\epsilon_x=m_\pi^2(T)x(T),
\ee
which tells that the approximate solution constructed by us obeys
Goldstone's theorem for the pions.  This feature of the optimized
perturbation theory was already emphasized in \cite{chiku} in the
context of the O(N) model.  We mention, however, that when the
symmetry breaking is realized by the appearance of two independent
order parameters, the application of PMS in the form of
Eq.~(\ref{Eq:M_T_gap}) cannot keep the mass of the other
pseudo-Goldstone boson, the kaon, at its tree--level expression.  This
means that the tree--level kaon mass does not satisfy the second
relation of Eq.~(\ref{ae1}) and Goldstone's theorem.  Had we chosen
for the mass--resummation the self-consistent treatment of the kaon
self-energy instead of pions, we would ensure that Goldstone's theorem
is fulfilled for the kaons. Both relations in Eq.~(\ref{ae1}) can be
fulfilled simultaneously only by resumming also one of the
higher-point functions of the theory in addition to the mass.

For the renormalization of $I(m_i,T)$ we wish to use such a
prescription, which allows to use further the parametrization of the
couplings realized with help of tree--level mass spectra. For this
reason we decided to omit all temperature independent finite
contributions from the tadpole and bubble integrals (the finite part
of the 1-loop $T=0$ corrections to the self-energy), retaining in
$I_R$ only the contributions from the part of the integrands
proportional to $n_B(\omega,T)$, which is the Bose-Einstein
distribution for a meson of energy $\omega$.  The explicit form of the
integral $I_R(m,T)$ with this prescription is the following:
\be
I_R(m,T)=\frac{1}{2\pi^2}\int_0^\infty 
dp\, p\, n_B(\sqrt{p^2+m^2}/T).
\ee
Now one can proceed to the solution of Eqs. (\ref{pigap}), (\ref{xeq}), 
(\ref{egy2}) for given $m_\pi, m_K$ when $T$ is varied. In the next
section we describe in detail how first order phase transitions were
detected and present the regions of the $(m_\pi - m_K)$--plane where
chiral symmetry restoring transitions take place
with increasing temperature. 

\section{The phase diagram in the $(m_\pi - m_K)$--plane}

In this section we present our results on the phase diagram in the
$(m_\pi - m_K)$-plane paying a special attention to the physical
point, the diagonal $m_\pi=m_K$ and the $m_\pi=0$ axis.  Investigating
the nature of the phase transition along the diagonal is important
because the result can be compared with lattice results
\cite{karsch01,karsch03} and also with previous results
\cite{Meyer-Ortmanns, Lenaghan00, lenaghan01, schmidt03}, obtained in
$L\sigma M$. Moreover due to the degeneracy in the particle spectrum,
the model is somewhat simpler on the diagonal, providing a good
testing ground for our approximation. The $m_\pi=0$ axis is relevant
because of the the presence of the tricritical point which separates
the region of first order phase transitions occurring for low values
of $m_K$ from the line of second order phase transitions.

Since we have two order parameters: $x$ (nonstrange) and $y$
(strange), we have to monitor both of them in order to decide the
nature of the phase transition. An interesting question arises whether
one can speak about two phase transitions, one related to the melting
of the nonstrange condensate and the other to the melting of the
strange condensate. The sign for a first order transition is the
appearance of three solutions for the equation of state
(\ref{eq:EoS_x}) and (\ref{eq:EoS_y}) below a given temperature,
corresponding to two minima and one maximum of the effective potential.
\begin{figure}[htbp]
\vspace*{-10pt}
\centering {\includegraphics[width=0.65\textwidth, keepaspectratio]{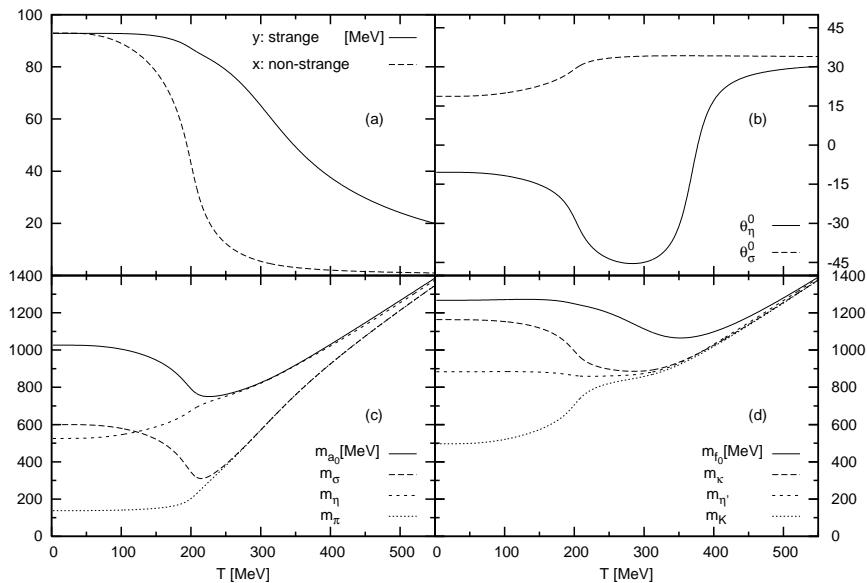}}
\caption{The temperature dependence  in the physical point of: 
(a)~the nonstrange (x) and strange 
(y) condensates; (b)~the pseudoscalar ($\theta_\eta$) and scalar 
($\theta_\sigma$) mixing angles (in the (0-8) basis); 
(c)~the mass of the chiral 
partners ($\pi,\sigma$) and ($a_0,\eta$); (d)~the mass off 
$f_0,\kappa,\eta',K$ mesons.} 
\label{fig:phys_vev_masses}
\end{figure}

In Fig.~\ref{fig:phys_vev_masses} we present our results on the
physical point, using alternative ,A2' with $m_\sigma(T=0)=600$
MeV. We preferred this one because alternative ,A1' gives
$m_\sigma(T=0)\approx900$ MeV, which is too high according to recent
phenomenological studies \cite{tornqvist02} and experiments
\cite{low_sigma}. The evolution of both
condensates at the physical point indicates a smooth crossover (see
Fig.~\ref{fig:phys_vev_masses} (a)), with a peak in the susceptibility
at around $T=210$ MeV for the nonstrange and $T=310$ MeV for the
strange case. The evolution of the strange condensate is much
slower. The restoration of the $SU(2)~\times~SU(2)$ symmetry can be
seen by observing the degeneracy between the $SU(2)$ chiral partners
($\pi,\sigma$) and ($a_0,\eta$), Fig.~\ref{fig:phys_vev_masses} (c).
We can observe the tendency of all the masses to converge at high
temperature. Note, however, the gap between the two sets of chiral
partners. This is the consequence of the $U_A(1)$ breaking determinant
term which enters with opposite sign in the expression of, for
example, $\pi$ and $a_0$ masses. This is insignificant only for very
small values of the strange condensate. The fact that, up to the
temperature shown in the figure, the $SU(3)$ chiral partners
($\pi,a_0$) and ($\eta,\sigma$) are not degenerate, indicates that the
restoration of the chiral symmetry is not completed in the strange
sector. We can also see in Fig.~\ref{fig:phys_vev_masses} (d) that the
variation of the strange condensate is reflected the strongest in the
mass of $f_0$ meson.

The evolution of the condensates and masses is nicely reflected also
by the temperature evolution of the mixing angles,
Fig.~\ref{fig:phys_vev_masses} (b). The pseudoscalar and scalar mixing
angles start at zero temperature at $\theta_\eta=-10.45^\circ$ and
$\theta_\sigma=18.7^\circ$ respectively, and they converge at high
temperature. Up to the temperature we studied, they do
not reach the ideal mixing angle $\arcsin(1/\sqrt{3})\simeq
35.264^\circ$, which means that $f_0$ and $\eta$ are not purely strange
mesons. In contrast to what was obtained in \cite{Lenaghan00,Costa}
the evolution of pseudoscalar mixing angle is nonmonotonic, it bends
down and then up as the temperature increases.

Next, we studied the phase boundary in the $m_\pi-m_K$-plane. As a
reference, we considered the case when each of the zero temperature
couplings of $L\sigma M$ has the fixed value calculated at the
physical point irrespective of the value of $m_\pi$ and $m_K$, except
for the external fields $\epsilon_x, \epsilon_y$, which follow the
variation of the $m_\pi$ and $m_K$ according to Eq.~(\ref{ae1}). For
$m_\sigma=900$ MeV we obtained nearly the same phase boundary as in
\cite{lenaghan01}. For $m_\sigma=600$ MeV no phase boundary was found
in \cite{lenaghan01} for $m_\pi>0$. With our method, the
phase boundary is present, but it is not compatible with the universality
requirement to have a first order transition in the neighborhood of
the origin.  In our view this represents an important argument for
allowing the variation of all couplings with $m_\pi$ and $m_K$.
\begin{figure}[htpb]
\vspace*{0pt}
\begin{minipage}[t]{0.49\textwidth}
\setlength{\abovecaptionskip}{0pt}
\centering{
\includegraphics[width=0.99\textwidth, keepaspectratio]{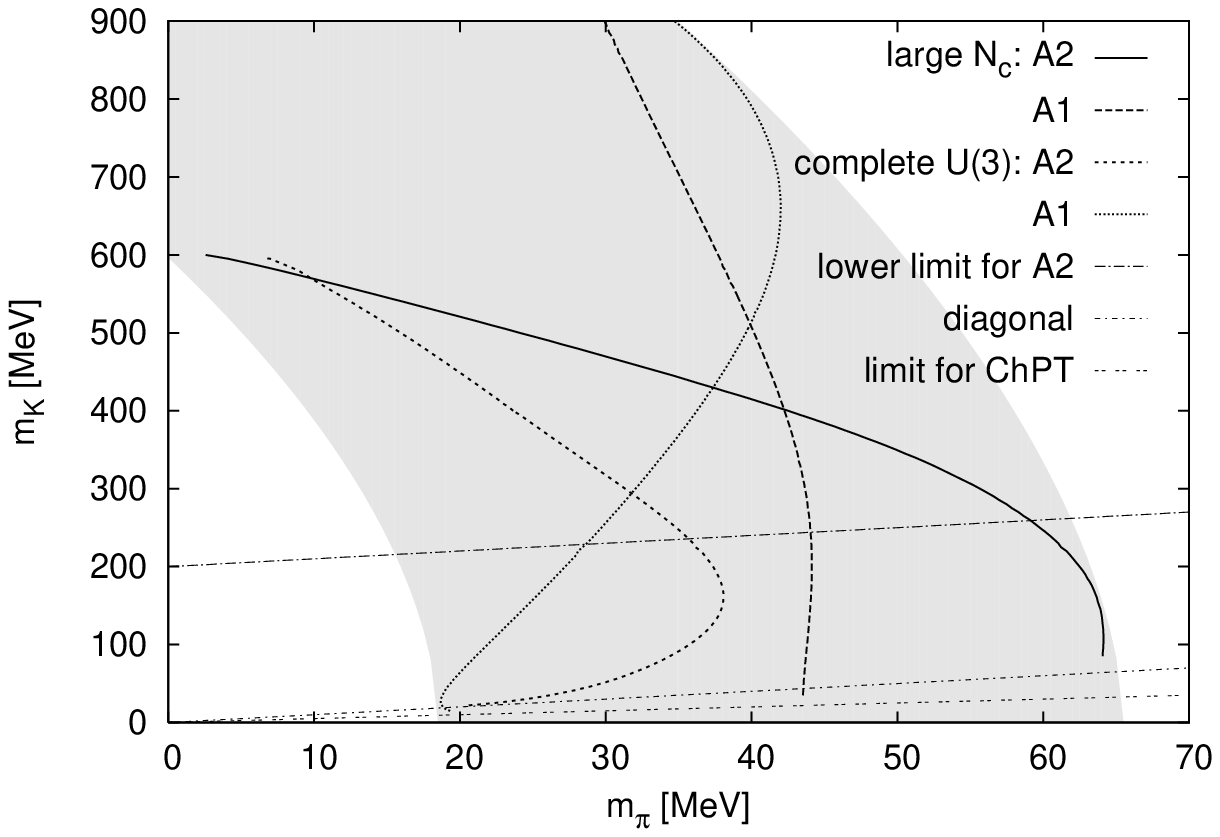}}
\caption{Phase boundary curves arising for $L\sigma M$
parametrizations compatible with $T=0$ ChPT. We present the boundary
curves for alternatives ,A1' and ,A2' using the complete U(3) ChPT and
the leading order large-$N_c$ ChPT at $\mathcal{O}(p^2)$. The exact
phase boundary curve is expected to lie in the light-grey shaded
region. Crossover (first order transition) takes places 
at the right (left ) of the shaded region.} 
\label{fig:PD_A1_A2}
\end{minipage}
\hfill
\begin{minipage}[t]{0.49\textwidth}
\setlength{\abovecaptionskip}{0pt}
\centering {\includegraphics[width=0.99\textwidth,keepaspectratio]{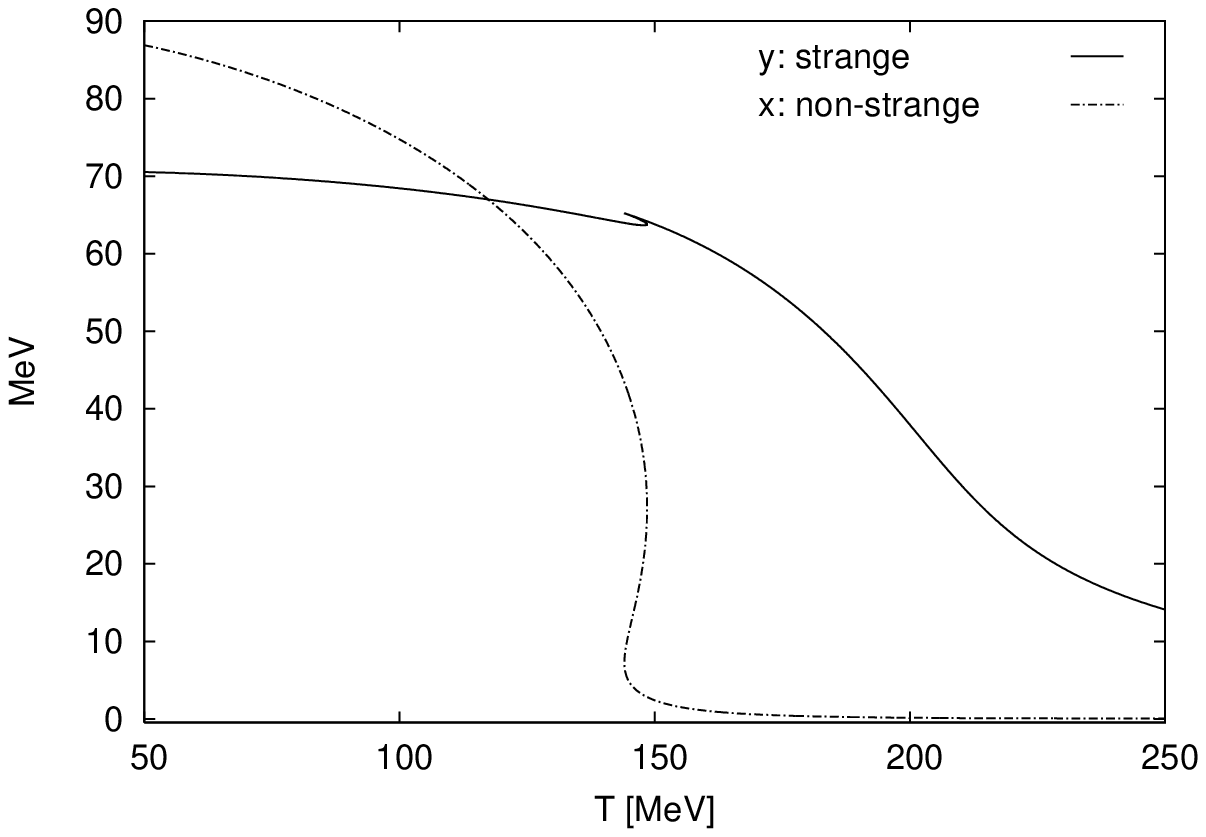}}
\caption{The temperature evolution of the nonstrange (x) and strange (y) 
condensates during a first order phase transition which takes place for 
$m_\pi=10$ MeV and $m_K=150$ MeV. The curves were obtained for alternative 
,A1' and the 
complete U(3) ChPT. Multivaluedness is observed in a given temperature 
range in both condensates.} 
\label{fig:x_y_T}
\vfill
\end{minipage}
\end{figure}

Fig.~\ref{fig:PD_A1_A2} presents the phase diagram obtained in the
case when all parameters are allowed to vary with $m_\pi$ and $m_K$.
Due to the uncertainties in the scalar sector and also due to
different approaches of the chiral perturbation theory (three-flavor
or large-$N_c$) we can give only a band indicative of the theoretical
uncertainties concerning the location of the real phase boundary of
the model. Note, however, that all variants give a first order
transition near the chiral limit, that is for small values of both the
pion and kaon masses. We see, that for any value of $m_K\leq 800 MeV$,
the critical value of the pion mass does not exceed $65$ MeV. Our
estimate for the phase boundary on the diagonal is
$m_\textrm{crit}(\textrm{diag})=40\pm 20$ MeV. In the figure also the
line is displayed below which we cannot trust the results of the
alternative ,A2', since its parameters diverge along the diagonal
$m_\pi=m_K.$

First order transitions are signalled by multivaluedness in the
temperature evolution of both the nonstrange and strange condensates,
see Fig.~\ref{fig:x_y_T}.  For large values of the kaon mass, we claim
that the phase transition is driven by the variation of the
nonstrange condensate, since the apparently different solutions of
the strange condensate are very close to each other, and all stay at
high values. Subsequent decrease of the strange condensate at higher
temperature displays only a crossover.  Along the phase boundary the
critical temperature reaches $T_c\approx 170$ MeV near the physical
kaon mass (in case ,A1' and using large $N_c$ ChPT), then drops to
$T_c=140$ MeV at both ends of the $m_K$ range shown in the figure,
which is due to the effect of the chiral logarithms.

We could not provide evidence for a tricritical point on the $m_\pi=0$
axis for any of the alternatives ,A1' and ,A2'. Alternative ,A1' seems
to predict it for such a high value of the kaon mass, where one can
not trust ChPT, while ,A2' does not work for $m_\pi=0$ because the
solution of EoSs leaves the physical region.

Finally, we discuss a feature of our approximate solution in the low
mass region which might be closely related to the problem of negative
squared masses.  It shows up the clearest along the diagonal,
$m_\pi=m_K$, where the most plausible expectation would be to have a
solution of EoS which satisfies, irrespective of the temperature, the
condition $\sigma_8=0$. For the alternative ,A1', (,A2' does not work
on the diagonal), it can be proved, using exclusively the tree--level
stability criteria $3f_1+f_2>0$, that going towards the origin below a
certain value of the Goldstone mass there is always a temperature
range in which one of the squared mass eigenstates in the mixing
scalar sector has negative eigenvalue. In this range we find in the
physical region only solutions with $\sigma_8\ne0$, that is the
'strange--nonstrange' symmetry is apparently broken in an
intermediate temperature range. It is not clear if this phase
corresponds to the absolute minimum of the free energy. This solution
is characterized by a large difference between the mass $m_K$
calculated from the second relation of Eq.~(\ref{ae1}) and the value
of the pion mass, which is the largest one.  In the mass region where
both solutions (with $\sigma_8=0$ and $\sigma_8\ne 0$) exist the
corresponding values of $\sigma_0$ are very close to
each other. Therefore, we expect that even in the case where we cannot
find the (true ?)  minimum corresponding to $\sigma_8=0$, a good
estimate of the position of the phase boundary on the diagonal is
provided by the $\sigma_8\ne0$ solution. The problem of negative
squared masses shows up also in the $T=0$ finite quantum correction
coming from the tadpole integrals, which were omitted in this work.

The above feature is a consequence of using tree--level expressions for
the propagator masses.  We certainly should have to go to higher--loop
order in the resummed perturbation theory, also to take into account
coupling resummations, for a complete resolution of the problem of
negative mass squares including the assessment of the solution with
$\sigma_8\ne 0$.

\section{Conclusions}

In this paper, we studied the phase boundary in the $(m_\pi -
m_K)$-plane allowing for the variation of all the parameters of the
linear sigma model with $m_\pi$ and $m_K$. We used for this another
low energy effective model, the chiral perturbation theory, which
being a perturbative expansion in momenta and in quark masses about
the chiral limit, provides, at each order of the momentum expansion,
analytical relations displaying the dependence of the decay constants
($f_\pi, f_K$) and masses of the $\eta$ and $\eta'$ on the value of
the pion and kaon masses.  One could expect that the linear sigma
model improved in this way will work reliably for small values of
$m_\pi$ and $m_K$. Using accurate formulas to continue from the
physical point, this approach could become an alternative to the
lattice which has difficulties in exploring this region when
information would be available on the variation of the mass of the
$\sigma$ or $f_0$ scalar mesons in the ($m_\pi -m_K)$-plane. The
origin of the theoretical uncertainty of our findings is the lack of
information on the scalar sector, which forces us to make assumptions.
Lattice results about the mass dependence in the scalar sector would
allow to reduce considerably the uncertainties of the parametrization
of the model.

The model was solved using a mass ressumation in the framework of the
optimized perturbation theory in order to resolve the negative squared
mass problem of the perturbation theory in the broken symmetry
phase. Unfortunately, resumming only one parameter, the mass, while
respecting the Goldstone's theorem for pions, violates Goldstone's
theorem for kaons. It also does not solve fully satisfactorily the
problem of negative mass squares in the whole mass--plane, since the
absolute minimum might be located in the $x-y$-plane slightly outside
the physical domain.  Resummation of another coupling is needed to
fulfill all requirements imposed by Goldstone's theorem.  A
possibility is to use the temperature variation of the coefficient of
the $U_A(1)$ violating term $g$. Motivated by lattice studies this
possibility was investigated in \cite{Costa}.

Taking into account all theoretical uncertainties, we could estimate a
band in the $(m_\pi - m_K)$-plane for the phase boundary.  Our
estimate for the boundary point on the diagonal is
$m_\textrm{crit}(\textrm{diag})=40\pm 20$ MeV, in nice agreement with
the latest effective model and lattice studies.

\appendix
\section{ \bf The $SU_L(3)~\times~SU_R(3)$ linear sigma model at tree--level} 

\subsection{\bf Mass eigenvalues, and mass matrices in the $0$-$8$ basis.
\label{App:eta_08}}

Using the inverse of the transformation (\ref{Otrans}), 
the mass matrix of $\eta$-s can be written in the more conventional 
$\eta_0$-$\eta_8$ basis:
\begin{eqnarray}
m^2_{\eta_{00}}&=&\frac{1}{3}(2m^2_{\eta_{xx}}+m^2_{\eta_{yy}}+
2\sqrt{2}\: m^2_{\eta_{xy}}),\\
m^2_{\eta_{88}}&=&\frac{1}{3}(2m^2_{\eta_{yy}}+m^2_{\eta_{xx}}-
2\sqrt{2}\:m^2_{\eta_{xy}}),\\
m^2_{\eta_{08}}&=&
\frac{1}{3}(\sqrt{2}(m^2_{\eta_{xx}}-m^2_{\eta_{yy}})
-m^2_{\eta_{xy}}),
\end{eqnarray}
The mass eigenvalues and mixing angle $\theta_\eta$ are the following:
{\allowdisplaybreaks
\bea
\displaystyle
&m^2_{\eta, \eta^{'}}=\frac{1}{2}(m_{\eta_{00}}^2+m_{\eta_{88}}^2\mp
\sqrt{(m_{\eta_{00}}^2-m_{\eta_{88}}^2)^2+4m_{\eta_{08}}^4}) \label{se},&\\
\label{Eq:mixing_angle}
&\displaystyle
\tan 2\theta_\eta=\frac{2m^2_{\eta_{08}}}{m^2_{\eta_{00}}-m^2_{\eta_{88}}},&
\eea}\noindent
where the ,-' sign refers to  $\eta$ and ,+' refers to $\eta^{'}$. These
expressions hold also for the mixing in the scalar sector, where the lower
mass eigenvalue is $m^2_\sigma$ and the higher is the squared mass of $f_0$.

\subsection {\bf The ,A2' alternative\label{App:A2}} 

For the scalar octet, there is a GMO mass relation
similar to the pseudoscalar sector, which to leading order reads:
\begin{equation}
4m^2_\kappa=m_{a_0}^2+3m^2_{\sigma_{88}}. \label{sgmo}
\end{equation} 
We characterize its accuracy by the following quantity: 
\begin{equation}
r:=\left. \frac{4m^2_\kappa-m_{a_0}^2}{3 m_{\sigma_{88}}^2}-1
\label{acc} \right.
. 
\end{equation} 
In the expression of $r$,  the mass squares $m_{a_0}^2$ and $m^2_\kappa$ are 
determined by $f_2, g, M^2$, which depend only on pseudoscalar mass 
squares: $m_\pi^2$, $m_K^2$, and $M_\eta^2=m_\eta^2+m_{\eta'}^2$.
In order to know
$m_{\sigma_{88}}$, we should have $f_1$ and $\mu_0^2$ separately. For this
purpose, in the physical point, we choose an $m_\sigma^\textrm{ph}$ to get
$r^\textrm{ph}$. We require that the accuracy of scalar GMO to be independent 
of $m_\pi$, $m_K$, that is 
$r(m_\pi,m_K,m_\sigma)=r(m_\sigma^\textrm{ph})=:r^\textrm{ph}$.
After this, we can
already determine $m_{\sigma_{88}}$ for arbitrary $m_K$, $m_\pi$ from
(\ref{acc}), and we can split $ M^2$ into $f_1$ and $\mu_0^2$. Using 
Table~\ref{Tab:masses} and Eqs.~(\ref{x})--(\ref{Mbar}) :
\begin{eqnarray}
\nonumber
f_1^{(A2)}&=&\frac{m_K^2((-64 f_K^3+104 f_\pi f_K^2-58 f_\pi^2 f_K+9 f_\pi^3) r
-12 (2 f_K-f_\pi) (f_K-f_\pi)^2)}
{32(3 f_\pi^2-8 f_\pi f_K+8 f_K^2)(f_K-f_\pi)^3 (r+1)}\\\nonumber&+&
\frac{m^2_\pi (r (16 f_\pi^3+8 f_K^3+24 f_\pi f_K^2-
39 f_\pi^2 f_K)+4 (f_\pi+2 f_K) (f_K-f_\pi)^2)}
{32(3 f_\pi^2-8 f_\pi f_K+8 f_K^2)(f_K-f_\pi)^3 (r+1)}\\&+&
\frac{M^2_\eta (f_K-f_\pi)^2 ( (2 f_K-f_\pi) r+ f_K-f_\pi)}
{4(3 f_\pi^2-8 f_\pi f_K+8 f_K^2)(f_K-f_\pi)^3 (r+1)},
\label{f_1_A2}\\
{\mu_0^2}^{\,(A2)}&=&f_1^{(A2)}(6f_\pi^2-8f_\pi f_K+8f_K^2)-{M^2}\label{mu_A2},
\end{eqnarray}
where $r^\textrm{ph} \approx -0.017$, when $m_\sigma^\textrm{ph}=600$ MeV.

\section{ \bf The U(3) C\MakeLowercase{h}PT } 
The elements of the mass matrix of $\eta$-s are defined in (\ref{forg}):
\begin{eqnarray}
m^2_{\eta_{88}}&=&D_{88}^{(0)}+d_{88}-(a_{88}D_{88}^{(0)}+
a_{08}D_{08}^{(0)})\,,  \label{88} \\ 
m^2_{\eta_{00}}&=&D_{00}^{(0)}+d_{00}-(a_{00}D_{00}^{(0)}+
a_{08}D_{08}^{(0)})\,,\\
m^2_{\eta_{08}}&=&D_{08}^{(0)}+d_{08}-\frac{1}{2}(a_{08}
(D_{00}^{(0)}+D_{88}^{(0)})+D_{08}^{(0)}(a_{00}+a_{88})) \label{08}  \,,
\end{eqnarray}
together with the mixing angle defined in Eq.~(\ref{Eq:mixing_angle}):
\be
\tan 2\theta_\eta=\frac{2D_{08}^{(0)}}{D_{00}^{(0)}-D_{88}^{(0)}}
\left[ 1-\frac{2d_{08}-a_{08}(D_{00}^{(0)}+D_{88}^{(0)})
-D_{08}^{(0)}(a_{00}+a_{88})}{2D_{08}^{(0)}}+
\frac{ d_{00}-d_{88}-a_{00}D_{00}^{(0)}+ a_{88}D_{88}^{(0)}
  } {D_{00}^{(0)}-D_{88}^{(0)}} \right]\,.
\ee
The relevant expressions of the $A$, $D$ matrices are given by 
\cite{Borasoy01}, \cite{Beisert01}:
{\allowdisplaybreaks
\begin{equation}
\begin{aligned}
 a_{00}& = \frac{1}{f^2}\frac{2}{3}A(2+q)(3L_4+L_5), \\
 a_{88}& = \frac{1}{f^2}(16A(2+q)L_4+\frac{16}{3}A(1+2q)L_5-2\mu_K ), \\
 a_{08}& = -\frac{16}{3\sqrt{2} }  \frac{1}{f^2}   A(q-1)L_5,
\end{aligned}\,\quad
\begin{aligned}
D_{00}^{(0)}&=-3v_0^{(2)}+\frac{2}{3}A(2+q), \\
D_{88}^{(0)}&=\frac{2}{3}A(1+2q), \\
D_{08}^{(0)}&=\frac{-2\sqrt{2}}{3}A(q-1), 
\end{aligned} \label{term1}
\end{equation}
\begin{equation}
\begin{aligned}
d_{88} &= \frac{1}{f^2}\frac{2A}{9}\left[{96 A} \left( (2+q)^2
  L_6+2(q-1)^2 L_7+(1+2q^2)L_8\right)-(1+8q)\mu_\eta-9\mu_\pi+6\mu_K\right],\\
d_{00} &=  -4A(2+q)(v_3^{(1)}-3v_2^{(2)})\\
&+\frac{1}{f^2}\frac{32A}{9}\left[A((2+q)^2(6L_6+2L_7)+6(2+q^2)L_8)
-\frac{1}{8}(\mu_\pi+6(1+q)\mu_K+(1+2q)\mu_\eta)\right],\\
d_{08} &= A(q-1)2\sqrt{2}v_3^{(1)}-\frac{1}{f^2}\frac{64\sqrt{2}}{3}
A^2(q^2-1)L_8. \label{term2} 
\end{aligned} 
\end{equation}
}\noindent
Therefore the elements of the mass matrix appearing in 
(\ref{88})--(\ref{08}) are:
{\allowdisplaybreaks
\begin{eqnarray}
m^2_{\eta_{88}}&=& \frac{2}{3}A(1+2q)+\frac{A}{f^2}
\left[ \frac{32}{3}A(1+2q^2)(2L_8-L_5)-\frac{32}{3}A(1+2q)(2+q)
L_4 \right.\no \\
&+& \left.\frac{128}{3}A(1-q)^2 L_7 +\frac{64}{3}A(2+q)^2L_6 - 
\frac{2}{9}(1+8q)\mu_\eta-2\mu_\pi+\frac{8}{3}(1+q)\mu_K\right] 
\label{etamass} \\
m^2_{\eta_{00}}&=&-3v_0^{(2)}+\frac{2}{3}A(2+q)\left[1-6v_3^{(1)}
+18v_2^{(2)}+\frac{1}{f^2}\left(24v_0^{(2)}(3L_4+L_5)\right)\right]\no \\
&-&\frac{A^2}{f^2}\left[\frac{16}{3}(2+q)^2(3L_4+L_5-6L_6-6L_7)\right]\no\\
&+&\frac{A}{f^2}\left[ \frac{64}{9}A(3(2+q^2)L_8-(1-q)^2L_5)
-\frac{4}{9}\mu_\pi+6(q+1)\mu_K+(2q+1)\mu_\eta)\right] \label{etamass2} \\
m^2_{\eta_{08}}&=&\frac{-2\sqrt{2}}{3} A(q-1)\left[1-3v_3^{(1)}
+\frac{1}{f^2}\left(12v_0^{(2)}L_5+16A(1+q)(2L_8-L_5)-
8A(2+q)L_4+\mu_K\right)\right]\label{etamass3}.
\end{eqnarray}
}\noindent
Carefully substituting the ${\cal O}(1/f^2)$
accurate expressions of $A$ and $q$  from (\ref{Acont}), (\ref{xcont})
into (\ref{etamass})-(\ref{etamass3})
we find for the variation of
$m_{\eta_{00}}^2, m_{\eta_{08}}^2, m_{\eta_{88}}^2$ in the 
($m_\pi - m_K$)--plane 
the following equations:
{\allowdisplaybreaks
\begin{eqnarray}
\label{Eq:m_eta88}
m_{\eta_{88}}^2&=&\frac{4m_K^2-m_\pi^2}{3}+\frac{1}{f^2}
\left[ \frac{8}{3}(\mu_K-\mu_\eta)m_K^2+\frac{2}{3}(\mu_\eta-\mu_\pi)
m_\pi^2+\frac{32}{3}(2L_8-L_5+4L_7)(m_\pi^2-m_K^2)^2\right. \no \\
&+&\left.\frac{32}{3}L_6(m_\pi^4-2m_K^4+m_K^2m_\pi^2)\right], \\
\label{Eq:m_eta00}
m_{\eta_{00}}^2&=&-3v_0^{(2)}+\frac{2m_K^2+m_\pi^2}{3}
\left(1-6v_3^{(1)}+18v_2^{(2)}+\frac{1}{f^2}24v_0^{(2)} 
(L_5+3L_4)\right)+\frac{1}{f^2}\left[\frac{4}{3}(-2\mu_K-\mu_\eta)
m_K^2\right. \nonumber \\
&+& \left. \frac{1}{3}(\mu_\eta-7\mu_\pi )m_\pi^2+ \frac{16}{3}
(2L_8-L_5)(m_\pi^2-m_K^2)^2+\frac{16}{3}L_7(m_\pi^2+2m_K^2)^2\right], \\
m_{\eta_{08}}^2&=&\frac{2\sqrt{2}}{3}\left\{(m_\pi^2-m_K^2)
\left[1-3v_3^{(1)}+\frac{1}{f^2}\left(12v_0^{(2)}L_5-8(2L_8-L_5)
(m_\pi^2-m_K^2)\right)\right]\right. \nonumber \\
&+&
\left.\frac{1}{f^2}\left[4(L_4-4 L_6)(m_\pi^4+m_K^2m_\pi^2-2m_K^4)+  
(\frac{1}{3}\mu_\eta-\mu_\pi+\mu_K) m_\pi^2
  +(\frac{2}{3}\mu_\eta-\mu_K) m_K^2\right] \right\}.
\end{eqnarray} 
}

\section{The tadpole coefficients in EoS \label{App:t_coeff}}

Below we list the nonzero three-point couplings, needed for the
evaluation of the tadpole contributions to EoS (see Eqs. (\ref{xeq}),
(\ref{egy2}))  in the $x-y$ basis (\ref{xybasis})

\be
\begin{pmatrix}  \alpha   & \, t^x_\alpha       & \, t^y_\alpha\\ 
                  \pi     & \, 2(2f_1+f_2)x     & \, 4f_1y+g  \\
                  K       & \,
                  2(2f_1+f_2)x-f_2\sqrt{2}y+\frac{1}{\sqrt 2}g &\,
                  -f_2\sqrt{2}x+ 4(f_1+f_2)y\\
                  \pi_x\pi_x &\, 2(2f_1+f_2)x &\, 4f_1y-g \\
                  \pi_y\pi_y &\, 4f_1x &\, 4(f_1+f_2)y \\
                  \pi_x\pi_y &\, -2g &\, 0\\
                  a_0 &\, 2(2f_1+3f_2)x &\, 4f_1y-g \\
                  \kappa &\, 2(2f_1+f_2)x+f_2\sqrt{2}y-\frac{1}{\sqrt
                  2}g &\, f_2\sqrt{2}x+4(f_1+f_2)y\\
                  \sigma_x\sigma_x &\, 6(2f_1+f_2)x &\, 4f_1y+g \\
                  \sigma_y\sigma_y &\, 4f_1x &\, 12(f_1+f_2)y \\
                  \sigma_x\sigma_y &\, 8f_1y+2g &\, 8f_1x \\
\end{pmatrix} 
.
\ee

The tadpole integrals are evaluated in the mass eigenbasis, therefore
in the pseudoscalar 
$x-y$ sector additional similarity transformations are needed
in order to arrive at the coefficients of the $\eta, \eta'$ tadpole
integrals. 
The elements of the $2 \times 2$ minor of $t^{x,y}$, which originally appears 
in the EoS, can be easily written in the mass eigenbasis. As an illustration
we take the pseudoscalar $(\eta, \eta')$ sector:
\be
\nonumber
t_{xx} G_{xx} + t_{yy} G_{yy} + t_{xy} G_{xy}=\tr\left[
\begin{pmatrix}
\displaystyle
t_{xx} & \frac{t_{xy}}{2}\\
\frac{t_{xy}}{2} & t_{yy}
\end{pmatrix}
\begin{pmatrix}
G_{xx} & G_{xy}\\
G_{xy} & G_{yy}
\end{pmatrix}\right]=
\tr\left[R(\theta)
\begin{pmatrix}
\displaystyle
t_{xx} & \frac{t_{xy}}{2}\\
\frac{t_{xy}}{2} & t_{yy}
\end{pmatrix}
R^T(\theta)
\begin{pmatrix}
G(m_{\eta'}) & 0\\
0 & G(m_\eta)
\end{pmatrix}\right],
\ee
where $G_{xx}$ is the $xx$ element of the $2\times 2$ propagator matrix,
$R(\theta)$ is an orthogonal transformation defined by 
$\tan 2\theta=2m^2_{\eta xy}/(m^2_{\eta xx}-m^2_{\eta yy})$ which
relates the $x,y$ and $\eta',\eta$ basis.

The diagonal $(\eta,\eta), (\eta',\eta')$ elements of the transformed
matrix are the coefficients of the corresponding physical propagators and can
be expressed as:
\be
t_{\eta',\eta}=\frac{1}{2}(t_{xx}+t_{yy}) \pm
\frac{(m^2_{\eta xx}-m^2_{\eta yy})(t_{xx}-t_{yy})+2m^2_{\eta xy}t_{xy}}
{2\sqrt{\left(m^2_{\eta xx}-m^2_{\eta yy}\right)^2+4m^4_{\eta xy}}}.
\ee

\section*{Acknowledgment}

The authors would like to thank A. Jakov\'ac and Z. Fodor for useful
discussions, and in particular, C. Schmidt, for providing 
information on his Ph.D-theses. We acknowledge the support of
Hungarian Research Fund (OTKA) under contract number T046129.

\end{document}